\documentclass{PoS}

\usepackage[margins, adjustmargins, final]{changes}
\usepackage{wrapfig}
\usepackage{xcolor}

\title{Enabling a new detection channel for beyond standard model physics with in-situ measurements of ice luminescence}

\ShortTitle{Luminescence of ice}

\author{
	The IceCube Collaboration\footnote{For collaboration list, see PoS(ICRC2019) 1177.}\\
	{\itshape \href{http://icecube.wisc.edu/collaboration/authors/icrc19_icecube}{http://icecube.wisc.edu/collaboration/authors/icrc19\_icecube}}\\
E-mail: \email{anna.pollmann@uni-wuppertal.de}
}

\abstract{

The IceCube neutrino observatory uses $1\,\mathrm{km}^{3}$ of the natural Antarctic ice near the geographic South Pole as optical detection medium. When charged particles, such as particles produced in neutrino interactions, pass through the ice with relativistic speed, Cherenkov light is emitted. This is detected by IceCube’s optical modules and from all these signals a particle signature is reconstructed.\\
A new kind of signature can be detected using light emission from luminescence. This detection channel enables searches for exotic particles (states) which do not emit Cherenkov light and currently cannot be probed by neutrino detectors. Luminescence light is induced by highly ionizing particles passing through matter due to excitation of surrounding atoms. This process is highly dependent on the ice structure, impurities, pressure and temperature which demands an in-situ measurement of the detector medium. \\
For the measurements at IceCube, a $1.7\,\mathrm{km}$ deep hole was used which {vertically} overlaps with the glacial ice layers found in the IceCube volume over a range of $350\,\mathrm{m}$. The experiment as well as the measurement results are presented. The impact {of the results, which enable new kind of} searches for new physics with neutrino telescopes, are discussed. \\

\vspace{4mm}
{\bfseries Corresponding author:}
\speaker{Anna Pollmann}$^{1}$\\
{$^{1}$ \itshape University of Wuppertal}

}

\FullConference{36th International Cosmic Ray Conference -ICRC2019-\\
		July 24th - August 1st, 2019\\
		Madison, WI, U.S.A.}

\begin{document}

\section{Introduction}\label{sec:intro}

{Luminescence} is the emission of light by a substance which is not resulting from heat. In particular \textit{radio-luminescence} \cite{Quickenden82} is light emission caused by ionizing radiation passing through a substance. To date the world largest particle detectors, i.e. {neutrino telescopes}, use water or ice as detection media. Reconstruction of particle properties relies on the optical detection of Cherenkov radiation emitted by charged particles at relativistic speeds passing through ice or water. However, slower moving particles, including particles proposed beyond the standard model, cannot be detected using this light emission channel. 

\begin{figure}
	\centering
	\includegraphics[width=.6\textwidth]{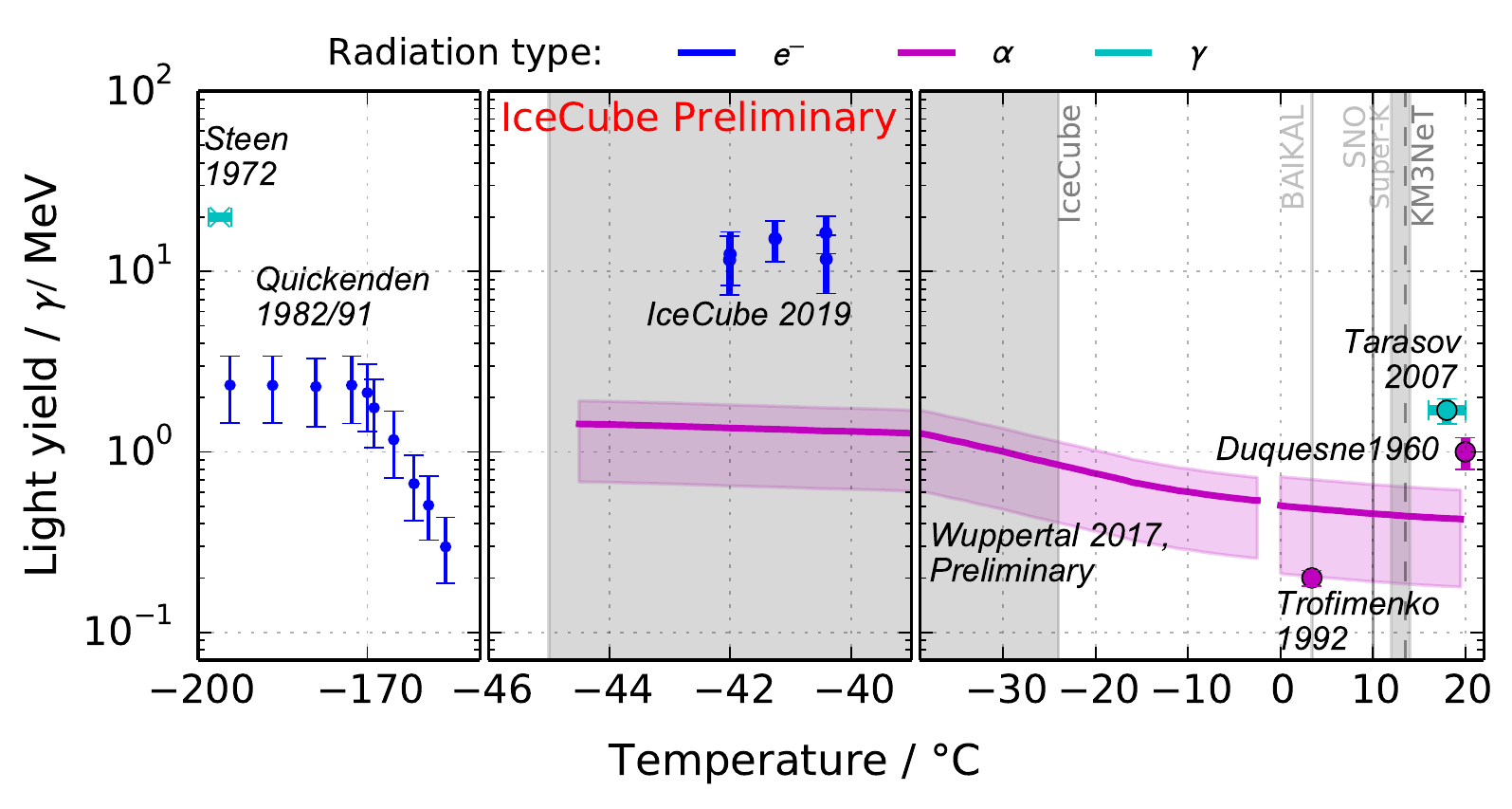}
	\caption{The result of this measurement (labeled as IceCube 2019) is shown in comparison to measured luminescence yields of cold ice, warm ice, and liquid water d by different kind of radiations, taken from Refs. \cite{Quickenden82, Duquesne60}. 
		Older measurements of cold ice luminescence are summarized in Ref. \cite{Quickenden82}. The water and ice temperatures of neutrino detectors are shown as vertical bands. 
		In addition to the values above, there is a recent measurement of water luminescence induced by $\alpha$-particles, protons, and carbon-ions which gives relative values only \cite{Yamamoto17}.	}
	\label{fig:ly_temp}
\end{figure}

The luminescence \textit{light yield} in water and ice is low and so far there are few measurements, as summarized in  Fig. \ref{fig:ly_temp}. The light yield depends on the energy deposition of the incident particle as well as its charge {due to} quenching. {Thus for highly ionizing particles
a measurable amount of luminescence light is expected}  \cite{Pollmann16}. The \textit{decay kinetics} of luminescence depend on the electronic transitions in which the photons are emitted. 

Laboratory measurements indicated that the luminescence light yield is highly dependent on solubles in water \cite{Pollmann2017b}. Therefore in-situ measurements of luminescence properties are required in order to use this light emission mechanism for the detection of particles in neutrino telescopes.
\textit{IceCube} is a cubic-kilometer neutrino detector installed in the ice at the geographic South Pole \cite{Aartsen:2016nxy} between depths of $1450\,\mathrm{m}$ and $2450\,\mathrm{m}$. The ice at South Pole is snow which was compacted due to pressure from new snow layers over time. Therefore it contains a significant amount of air.

\section{Setup}\label{sec:setup}

{About a kilometer from the boundary of the IceCube array,} a $1750\,\mathrm{m}$ deep borehole was drilled by the \textit{SPICEcore project} \cite{SpiceCore2014} from 2014 to 2016. The borehole has a diameter of $126.8\,\mathrm{mm}$ and is filled with {an antifreeze liquid}\footnote{Estisol-140}. A significant tilt in the ice layers \added{means}\deleted{induces} that the {vertical} overlap between the ice probed by IceCube and SPICEcore is effectively $\sim 350\,\mathrm{m}$. The temperature in the borehole reaches from about $-50^{\circ}\mathrm{C}$ close to the surface up to approximately $-30^{\circ}\mathrm{C}$ at the lower end.
{In the 2018/2019 season two other optical experiments, than the device described below, were deployed in the SPICEcore borehole: 
the Camera System and the UV-Logger \cite{Camera}.}

\begin{figure}
	\centering
	\includegraphics[width=1.0\textwidth]{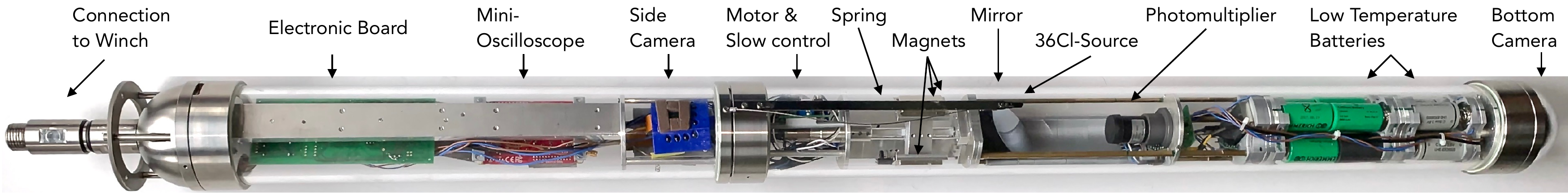}
	\caption{Photograph of the luminescence logger (turned by $90^{\circ}$).
	}
   \label{fig:logger}
\end{figure}

A logging device, called \textit{Luminescence Logger}, was built in order to measure the luminescence yield and decay kinetics in the SPICEcore hole, see Fig. \ref{fig:logger}. 
A radioactive source is attached at the end of a flat spring outside of a pressure vessel made out of quartz glass. 
{The isotope $^{36}\mathrm{Cl}$, contained in a titanium housing, was chosen because it emits $\beta$-radiation with an endpoint at $540\,\mathrm{keV}$ outside the titanium.}
The source can be pushed against the borehole wall with the help of magnets.

\added{Behind}\deleted{ the height of} the source there is a parabolic mirror in the vessel which reflects photons emitted close to the source onto a photomultiplier\footnote{Hamatsu R1924P, 1 inch, with magnetic shielding}. The photomultiplier is read out with a FPGA based oscilloscope\footnote{RedPitaya, STEMLab 125-14} in the logger 
which can record timestamps and waveforms up to $40\,\mathrm{kHz}$ \added{and a trigger}\deleted{with a} rate \deleted{accuracy}\added{with an accurancy} of 95.5\%. A deadtime of about $800\,\mathrm{ns}$ was deduced from measurements. The oscilloscope sends monitoring information to a computer at the surface above the borehole using RS-485. Power for most devices is provided by the winch cable from surface.  The motor, which drives the magnet rotor, is powered by a set of batteries. A logic circuit is implemented which moves the rotor into save position (spring attached) when connection from surface to the logger is lost. 


%

The logger was deployed twice into the SPICEcore borehole {to a depth of $1560\,\mathrm{m}$ in the austral summer 2018/2019.} 
During the first run the source was not attached in order to test for mechanical safety and take background measurements at 5 different depths. 
The measured temperature in the borehole increases with increasing depth from $-48^{\circ}\mathrm{C}$ to $-36^{\circ}\mathrm{C}$ at the lowest measurement point.

After all tests were finished successfully, the source was attached for the signal measurement at three different positions around {depths of} $1399\,\mathrm{m}$, $ 1474\,\mathrm{m}$, and $1559\,\mathrm{m}$. 
In every position two measurements were taken about $75\,\mathrm{cm}$ apart in order to account for local effects in the ice. 
In addition a measurement was taken when the radioactive source was not attached to the ice in order to measure the light induced in Estisol {(labeled as measurement \textit{2/2c} in the figures)}.  

\begin{figure}
	\centering
	\includegraphics[width=.49\textwidth]{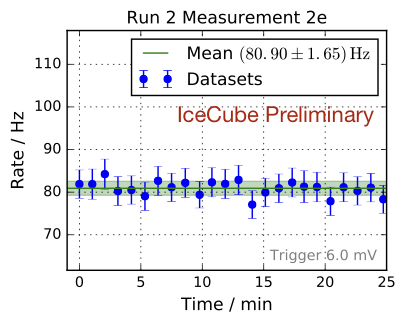}
	\includegraphics[width=.49\textwidth]{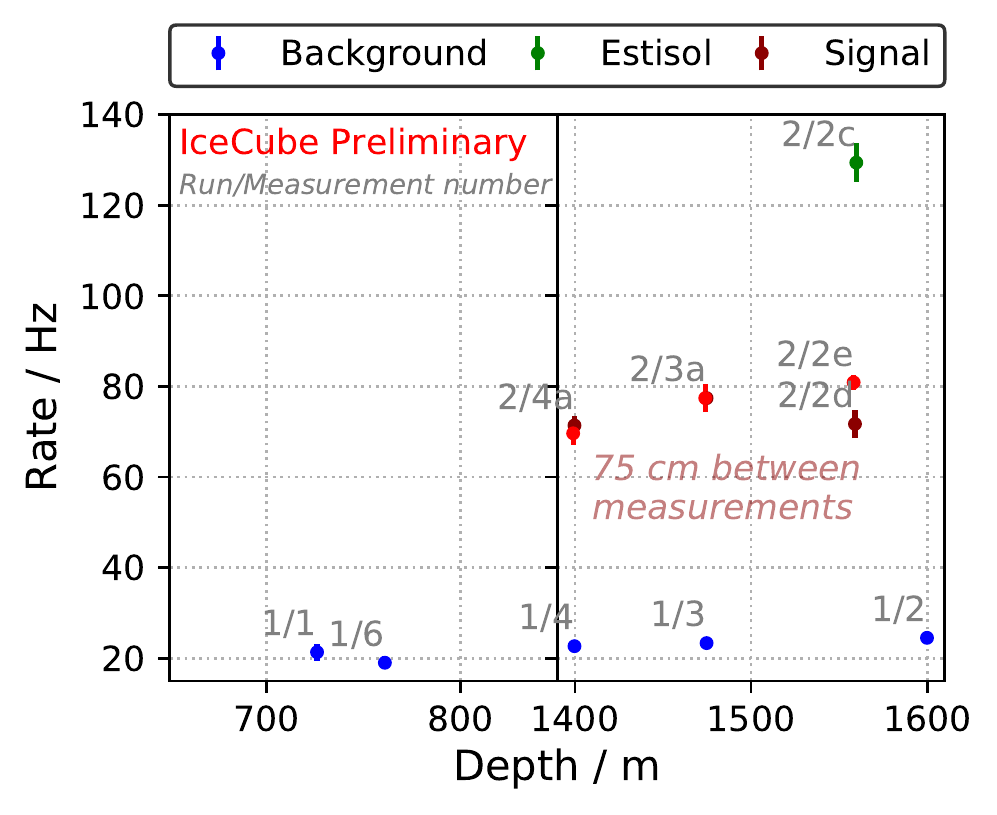}
	\caption{\textbf{Left:} Rate over time for the signal measurement at $1558.25\,\mathrm{m}$ depth. The average rate, which was used for further analysis, is shown in the legend. Error bars show statistical uncertainties.
	\textbf{Right:} Comparison of the average rates of all taken measurements. Different colors denote different experimental setups, see description in the text. The labels of measurements {show the order of the measurements and whether they were taken in the first or second run}. Errorbars show the standard deviation of the average rate. Signal measurements were taken twice per depth about $75\,\mathrm{cm}$ apart.
	}
	\label{fig:rate}
\end{figure}

For each pulse, which exceeded an adjustable threshold, the timestamp and a waveform of $120\,\mathrm{ns}$ with $250\,\mathrm{MHz}$ sampling rate and $0.12\,\mathrm{mV}$ resolution was taken. Offline all datasets were discarded for which the logs showed potential issues, such as e.g. a broken generator or communication issues. Also \added{a fixed}\deleted{an equal} discrimination threshold of $6\,\mathrm{mV}$ was added offline.  The rate over time for an individual measurement is shown in Fig. \ref{fig:rate} (left). All other measurements were similarly stable in rate. 
Therefore an average rate was calculated for each measurement type and depth,  shown in Fig. \ref{fig:rate} (right).

\section{Analysis and results}\label{sec:results}


In order to obtain the light yield of luminescence in ice, the optical setup was modeled with a custom ray tracing program. 
The input into the program are simulated electrons \cite{GEANT} from the radioactive source. The electrons induce emission of Cherenkov light and luminescence light in both, Estisol and ice. The average numbers of photons per emission type is calculated from the electrons' speeds and energy losses. The actual number of photons is then drawn from a Poisson distribution around the {aforementioned} value. The starting positions of the photons are spread randomly over the electrons' track parts. The starting direction of the photon in relation to the electron is either the Cherenkov angle or drawn from an isotropic distribution for luminescence. The photon track lengths is drawn from an exponential with the attenuation length in ice. The photons are then propagated until they are absorbed while accounting for scattering, refraction and reflection. The number of photons from different emission types reaching the photomultiplier is counted and can be converted into an estimated rate using the emission rate of the radioactive source. 

\begin{figure}
	\centering
		\includegraphics[width=.42\textwidth]{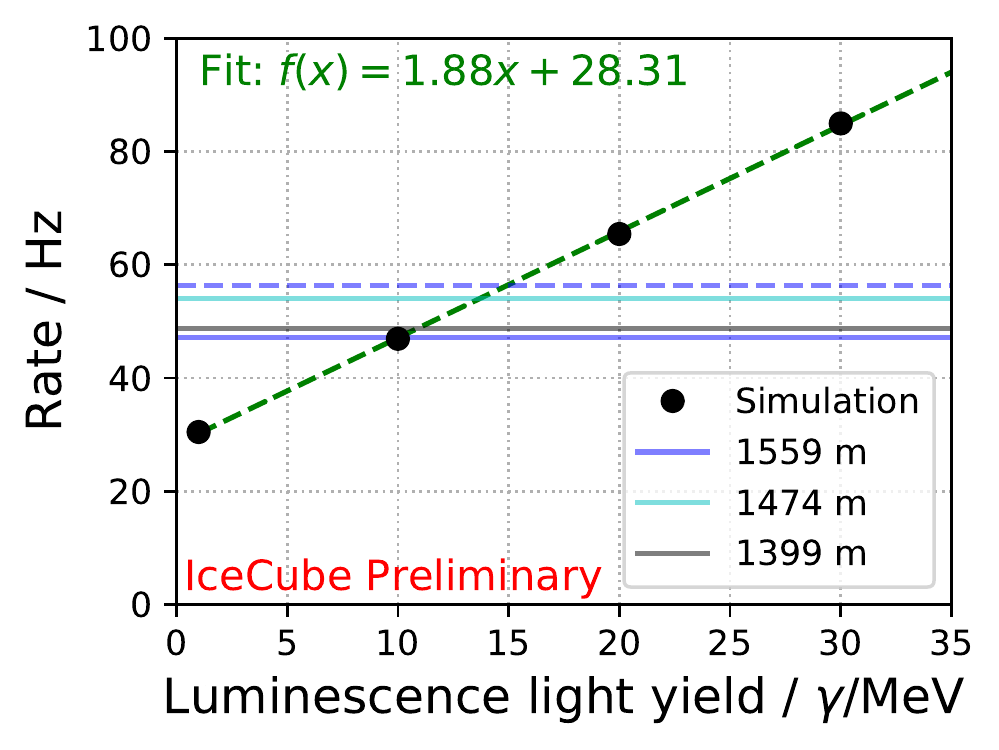}
		\includegraphics[width=.32\textwidth]{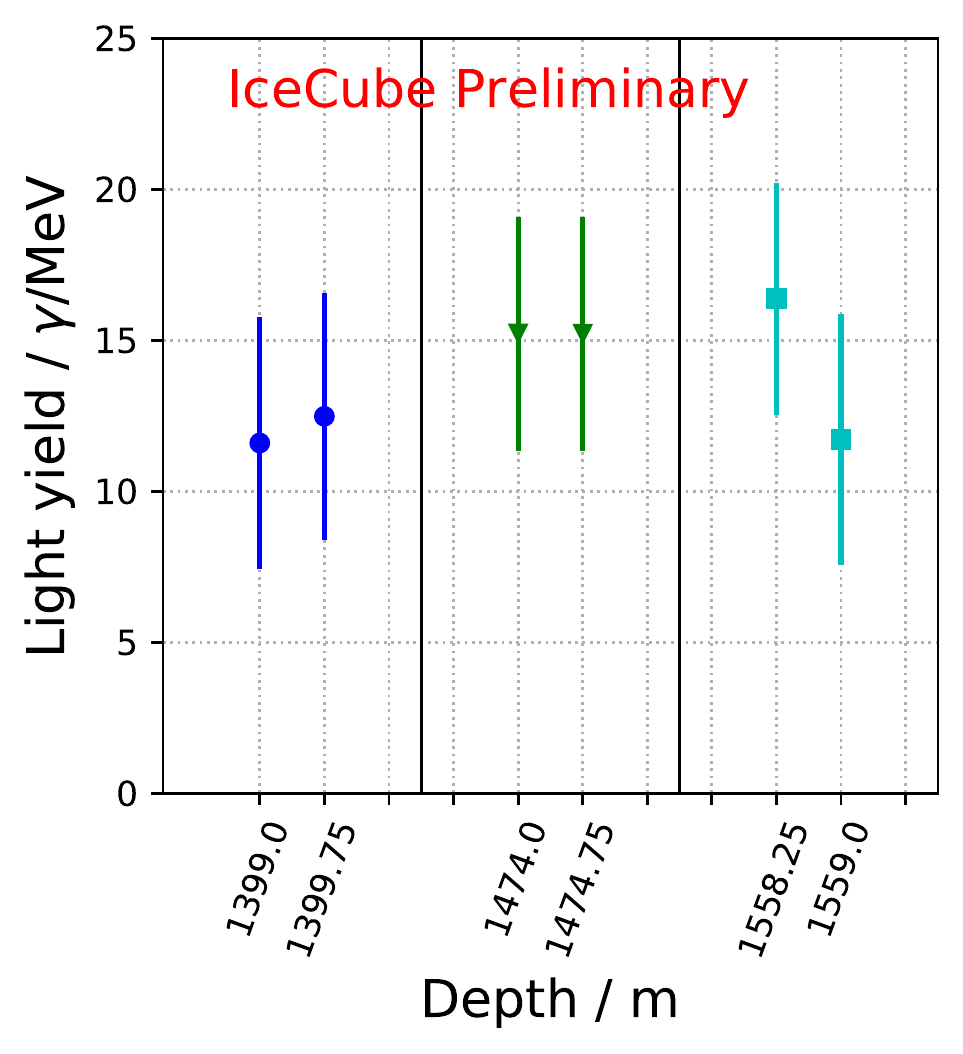}
	\caption{\textbf{Left:} Estimation of the luminescence light yield by comparing simulated rates (black dots and fit in dashed green) with measured rates (horizontal lines). 
	\textbf{Right:} Comparison of the luminescence measurement of ice measured in different depth and at different temperatures in South Pole ice. The contributing uncertainties are described in the text.}
	\label{fig:results}
\end{figure}

The rate of dark noise as well as photons emitted in Estisol due to luminescence and Cherenkov light, are assumed to be approximately constant. The light yield of ice luminescence is varied in the simulation as well as the average distance of the source to the ice. The predicted rates are compared with the measured rates which gives a range of possible values for the luminescence yield of ice per measurement. Uncertainties from the mirror reflectivity ($\pm10\%$), photomultiplier quantum efficiency ($\pm10\%$), source emission rate ($-19\%,+11\%$), scattering length and absorption length ($\pm13\%$) in ice are investigated and included in the final result which is shown in Fig. \ref{fig:results}. Since a quenching ratio of electrons to alphas of about 10 is expected \cite{Quickenden82}, the measured light yield is only slightly higher than expected from the laboratory measurement shown in Fig. \ref{fig:ly_temp} for comparison.

The electronic transitions emitting luminescence photons are expected to have different decay times. An exponential decrease of photon counts after an excitation is expected. In this measurement only single photons are detected, hence no initial pulse can be identified. However, choosing a random pulse as being close in time to the initial excitation and considering all the time length between this and all subsequent pulses (up to $2\,\mathrm{ms}$) still leads to the same shape of distribution. 

\begin{figure}
	\centering
	
	\includegraphics[width=.42\textwidth]{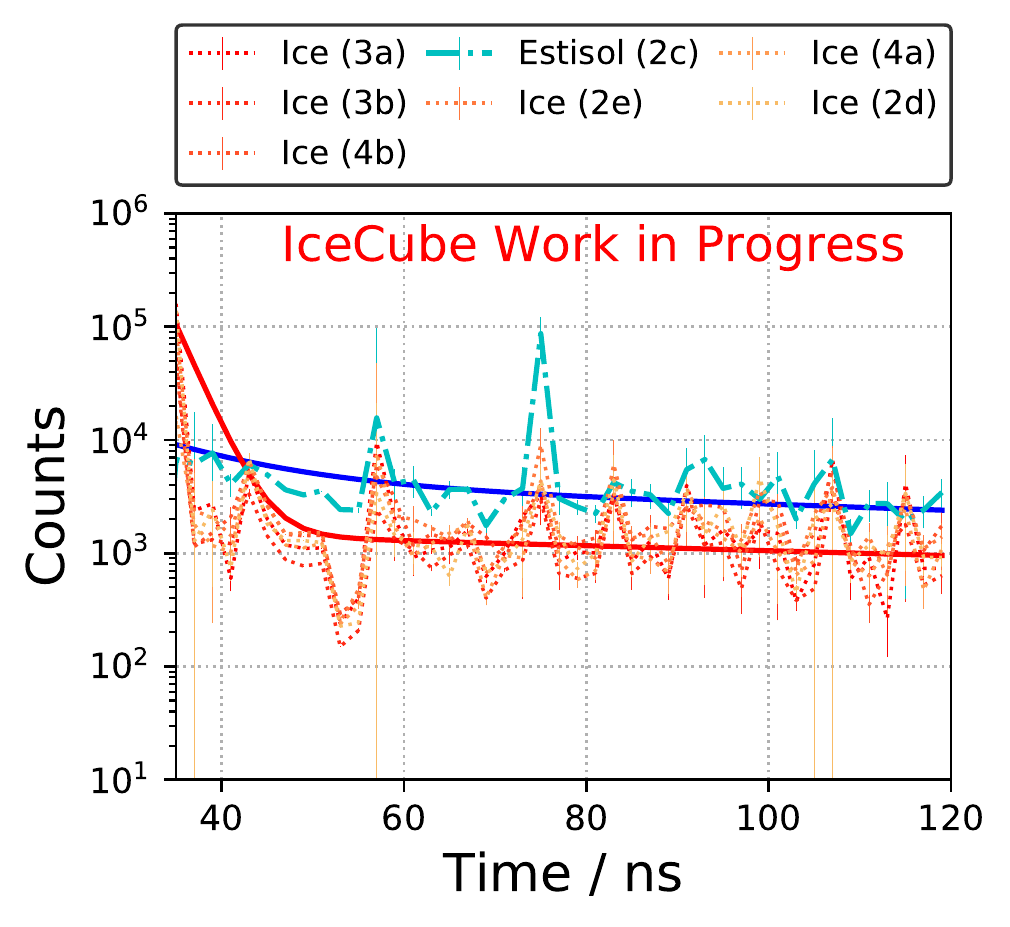}
	\includegraphics[width=.42\textwidth]{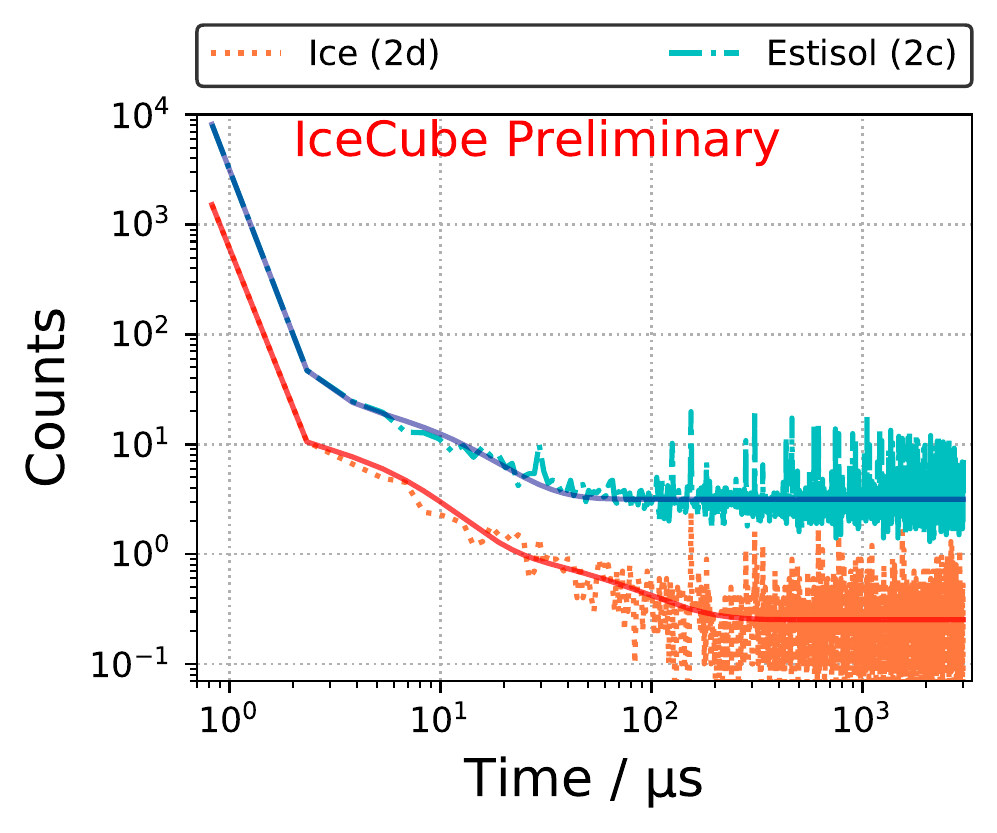}
	\caption{Distributions of the time differences of pulses following a random initial pulse for short time scales (left) and long times scales (right). The labels of each measurement are equal to the labels in Fig. \ref{fig:rate} (right). {Error bars on the left plot show propagated statistical uncertainties}. For clarity only one signal measurement is shown on the right plot. }
	\label{fig:results_time}
\end{figure}

The datasets of the measurements were cleaned in the same way as for the analysis of the light yield. For short time scales the waveforms are analyzed in order to identify pulses, see Fig. \ref{fig:results_time}.
For longer time scales the trigger time stamps are used. Since there is a dead time of about $800\,\mathrm{ns}$, there is no information available between $120$ and $800\,\mathrm{ns}$. The contribution of Estisol luminescence was found to be negligible in the light yield analysis and is therefore not subtracted. 

In order to account for PMT effects and electronic ringing a sample distribution is obtained for short time scales from the dark noise measurement and the signal distributions are corrected for this shape. 
The short and long time distributions of pulses is fit with the minimum number of exponential distributions which describes the distributions from $40\,\mathrm{ns}$ to about $200\,\mathrm{\mu s}$ and minimize the $\chi^2$ of the fit. Four decay times could be \added{preliminarily} identified for ice luminescence:  
$(2.44\pm 0.21)\,\mathrm{ns}$,   
$(189.6\pm 29.9)\,\mathrm{ns}$, 
 $(5.03\pm 0.06)\,\mathrm{\mu s}$, and
  $(56.10\pm 6.26)\,\mathrm{\mu s}$.
There \added{are}\deleted{is} no \added{previous}\deleted{comparable} measurements to which these values can be \added{compared}\deleted{matched}.

\section{Application of luminescence}\label{sec:appl}

\begin{figure}
	\centering
	\includegraphics[width=.95\textwidth]{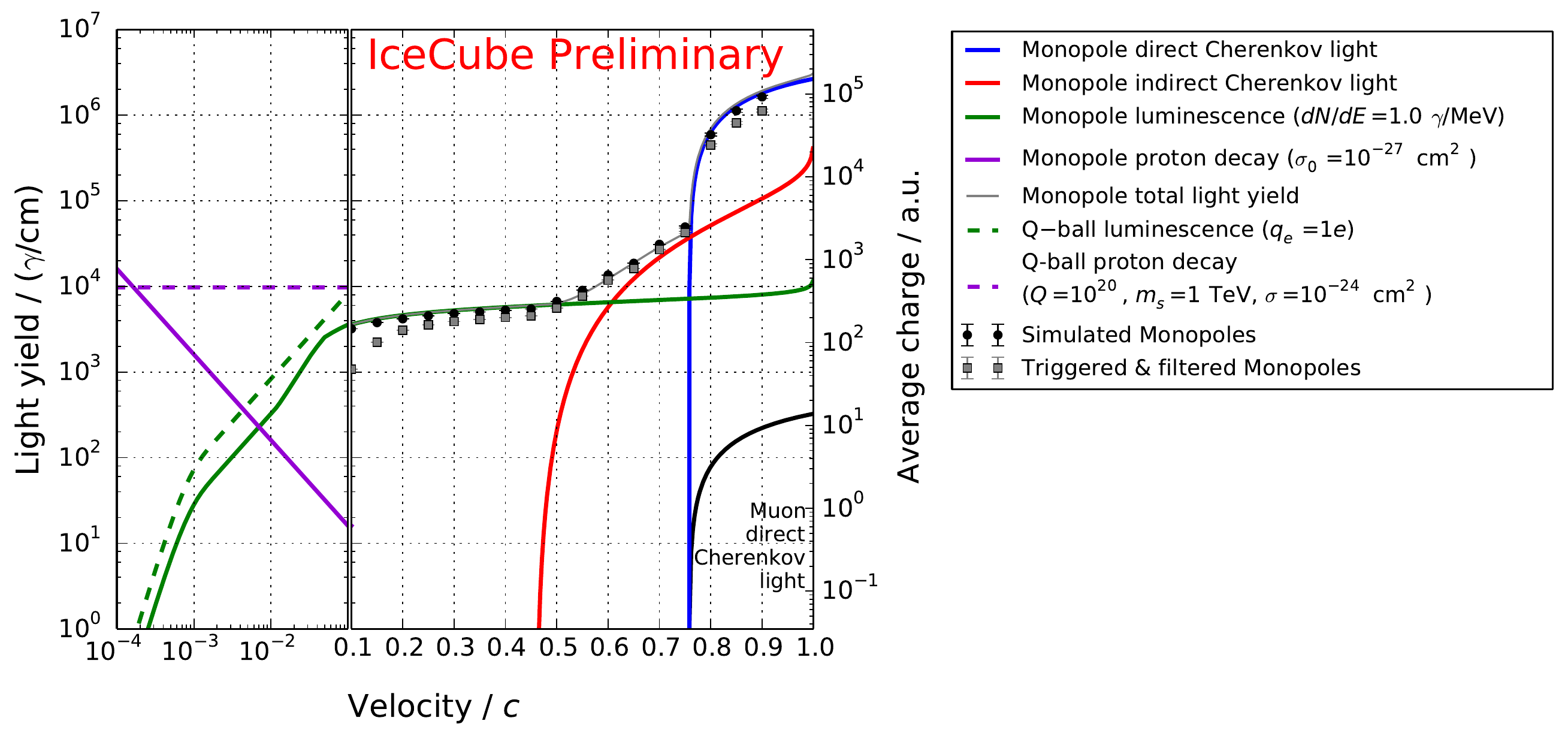}
	\caption{Light yield of Cherenkov light, luminescence and nucleon decay  from magnetic monopoles in ice (solid colored lines) in comparison to the light yield of Q-balls (dashed lines) for luminescence  or nucleon decay and muons emitting Cherenkov light (black line). Simulation of different properties \added{are}\deleted{is} shown for magnetic monopoles as well as the trigger efficiency on those particles for IceCube.
	}
	\label{fig:nphotons}
\end{figure}

Two of the particles which can only be searched for with neutrino telescopes using luminescence in ice are charged Q-balls as well as slow and mildly relativistic magnetic monopoles.
Q-balls are predicted in supersymmetric theories \cite{Kusenko97} to be coherent states of squarks and sleptons and {they are} candidates for dark matter. Neutral Q-balls can catalyze nucleon decay with the KKST process \cite{Kusenko98}. These can be searched for using similar approaches as searches for slow magnetic monopoles catalyzing nucleon decay via the Rubakov-Callan effect \cite{Rubakov88}.  Charged Q-balls emit luminescence light and the light yield of this process, shown in Fig. \ref{fig:nphotons}, is obtained by convoluting the energy loss with the luminescence efficiency obtained above. 

Magnetic monopoles are predicted in all GUT theories to be massive particles which carry at least one magnetic charge (see a summary in Ref. \cite{Lauber17}). At low speeds they can catalyse nucleon decay in some theories. At mildly and highly relativistic speeds they emit (indirect) Cherenkov light. In between those parameter ranges, i.e. below $0.5\,c${, with $c$ being the speed of light in vacuum,} and for monopoles which do not catalyse nucleon decay, only luminescence can be used for detection, see Fig. \ref{fig:nphotons}. 

\begin{figure}
	\centering
	\includegraphics[width=.69\textwidth]{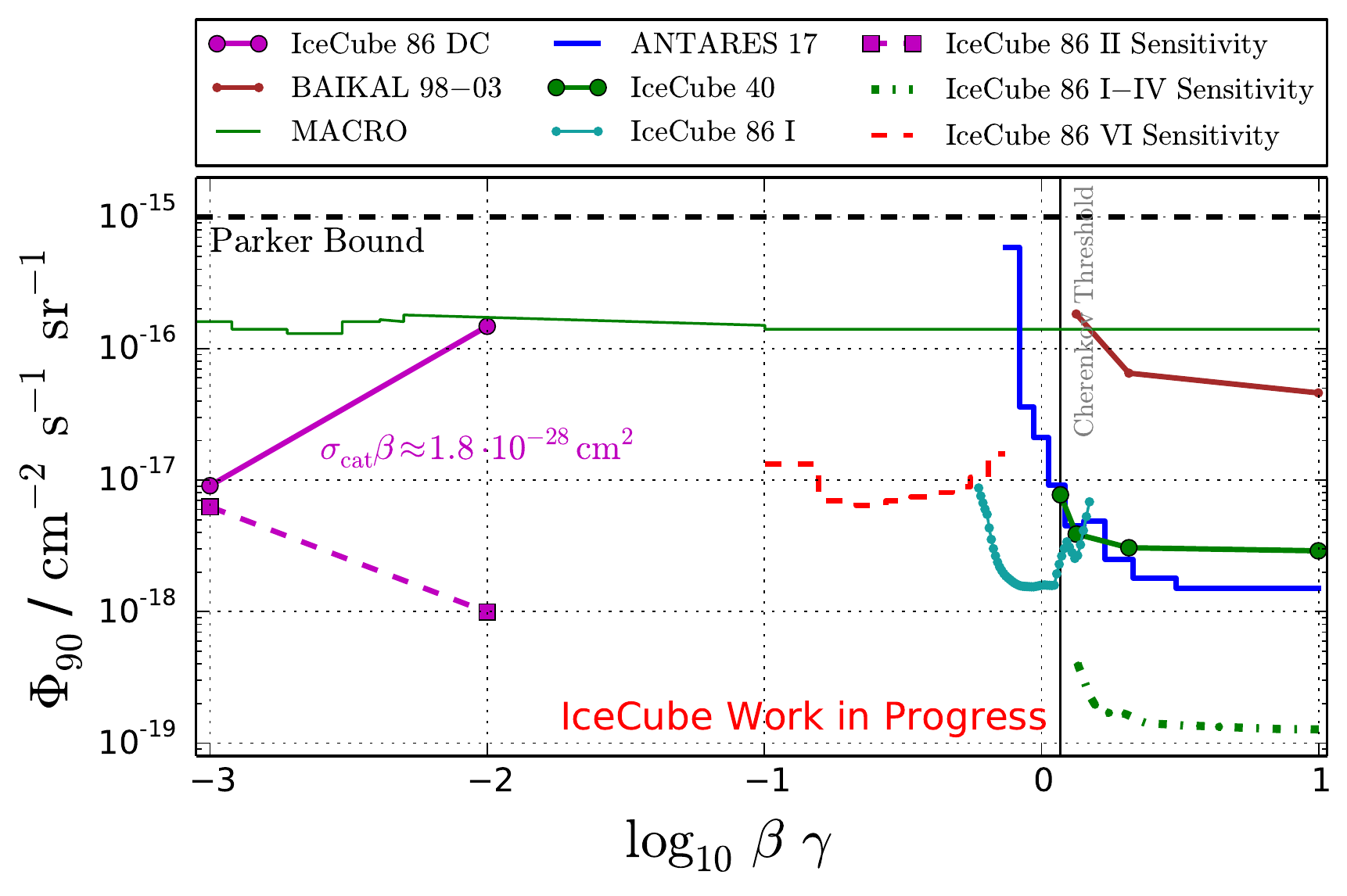}
	\caption{ Limits and sensitivities of recent searches for magnetic monopoles \cite{Schoenen14} in comparison to the first search using luminescence of ice as detection chanel (red dashed line).}
	\label{fig:sensitivity}
\end{figure}

\deleted{There is} The first analysis \deleted{ongoing} using luminescence light to search for low relativistic magnetic monopoles \added{is being done} in IceCube \cite{Lauber17}.  {The events signature of these monopoles is homogeneous over the track and comparably dim. 
The background consists mostly of muons originating from air showers which are coincident in time.
The slower \added{the speed of} an incident monopole the less light is recorded since the triggers are optimized for faster particles, as shown in Fig. \ref{fig:nphotons}. } A dedicated filter was written in order to select matching signatures for further analysis. On this data several \deleted{straight} cuts \added{on reconstructed variables} are applied in order to select the considered speeds and remove events with few hits only. Afterwards a machine learning algorithm, a gradient boosted decision tree (BDT) \cite{XGBoost}, is used to separate background and signal. Variables used in the BDT are e.g. the timelength of the event, the length of the track, the (homogenioty of the) brightness, and its location in the detector. Finally a resampling technique \cite{Pollmann2017c} is used to account for low statistics in low energetic and coincident background events since IceCube analyses are optimized on simulation. Finally a cut on the output of the BDT is optimized to obtain the highest sensitivity. It reduces the background rate by several orders of magnitude.
The estimated sensitivity \added{to magnetic monopoles} exceeds previous exclusion limits by an order of magnitude, see Fig.\,\,\ref{fig:sensitivity}.

\section{Discussion and outlook}\label{sec:outlook}

For the first time luminescence yield and decay kinetics were measured in the detection medium of a neutrino telescope. The measurement results are already in use by two analyses and lead to a highly competitive sensitivity for searches \added{for physics} beyond the standard model. 

A second deployment of the luminescence logger is planned in which \added{luminescence} measurements will be taken at more depths with smaller uncertainties. In addition to the light yield and the decay time, a rough spectrum will be measured in order to identify the exact processes which lead to the emission of luminescence light in ice.
\\

\textbf{Acknowledgements: } 
The authors would like to thank the SPICEcore collaboration for providing the borehole, the US Ice Drilling Program, \added{the Antarctic Support Contractor and the National Science Foundation} for providing the equipment to perform the described measurement \added{and for their support at South Pole}.


\bibliographystyle{ICRC}
\bibliography{references}

\end{document}